\documentstyle[12pt,epsf,epsfig]{article}
\newfont{\thiplo}{msbm10 scaled\magstep 2}
\newfont{\gothic}{eufb10 scaled\magstep 2}
\newfont{\unc}{eurb10} 
\newskip\humongous \humongous=0pt plus 1000pt minus 1000pt
\def\caja{\mathsurround=0pt}
\def\eqalign#1{\,\vcenter{\openup1\jot \caja
        \ialign{\strut \hfil$\displaystyle{##}$&$
        \displaystyle{{}##}$\hfil\crcr#1\crcr}}\,}
\newif\ifdtup

\def\eqright #1\cr{\noalign{\hfill$\displaystyle{{}#1}$}}
\def\eqleft #1\cr{\noalign{\noindent$\displaystyle{{}#1}$\hfill}}

\def\oldreffmt#1{\rlap{[#1]} \hbox to 2\parindent{}}

\def\figfmt#1{\rlap{Figure {#1}} \hbox to 1in{}}

\def\sectioneq{\def\theequation{\thesection.\arabic{equation}}{\let
\holdsection=\section\def\section{\setcounter{equation}{0}\holdsection}}}%

\newcounter{holdequation}

\def\auto{\eqno(\refstepcounter{equation}\theequation)}
\def\begineq #1\endeq{$$ \refstepcounter{equation}\eqalign{#1}\eqno
	(\theequation) $$}
\def\contlimit{\,{\hbox{$\longrightarrow$}\kern-1.8em\lower1ex
\hbox{${\scriptstyle (a\rightarrow0)}$}}\,}
\def\centeron#1#2{{\setbox0=\hbox{#1}\setbox1=\hbox{#2}\ifdim
\wd1>\wd0\kern.5\wd1\kern-.5\wd0\fi
\copy0\kern-.5\wd0\kern-.5\wd1\copy1\ifdim\wd0>\wd1
\kern.5\wd0\kern-.5\wd1\fi}}
\def\centerover#1#2{\centeron{#1}{\setbox0=\hbox{#1}\setbox
1=\hbox{#2}\raise\ht0\hbox{\raise\dp1\hbox{\copy1}}}}
\def\centerunder#1#2{\centeron{#1}{\setbox0=\hbox{#1}\setbox
1=\hbox{#2}\lower\dp0\hbox{\lower\ht1\hbox{\copy1}}}}
\def\lsim{\;\centeron{\raise.35ex\hbox{$<$}}{\lower.65ex\hbox
{$\sim$}}\;}
\def\gsim{\;\centeron{\raise.35ex\hbox{$>$}}{\lower.65ex\hbox
{$\sim$}}\;}
\def\st#1{\centeron{$#1$}{$/$}}

\def\super#1{\ifmmode \hbox{\textsuper{#1}}\else\textsuper{#1}\fi}
\def\textsuper#1{\newcount\holdspacefactor\holdspacefactor=\spacefactor
$^{#1}$\spacefactor=\holdspacefactor}

\def\getcite#1,{\advance\citenumber by1
\def\getcitearg{#1}\def\lastarg{@}
\ifnum\citenumber=1
\ref{#1}\let\next=\getcite\else\ifx\getcitearg\lastarg\let\next=\relax
\else ,\ref{#1}\let\next=\getcite\fi\fi\next}

\def\pom{{\rm P\kern -0.53em\llap I\,}}
\def\spom{{\rm P\kern -0.36em\llap \small I\,}}
\def\sspom{{\rm P\kern -0.33em\llap \footnotesize I\,}}

\relax
\def\contlimit{\,{\hbox{$\longrightarrow$}\kern-1.8em\lower1ex
\hbox{${\scriptstyle (a\rightarrow0)}$}}\,}
\def\upon #1/#2 {{\textstyle{#1\over #2}}}
\relax
\renewcommand{\thefootnote}{\fnsymbol{footnote}} 

\def\mainhead#1{\setcounter{equation}{0}\addtocounter{section}{1}
  \vbox{\begin{center}\large\bf #1\end{center}}\nobreak\par}

\def\til#1{\centeron{\hbox{$#1$}}{\lower 2ex\hbox{$\char'176$}}}
\def\tild#1{\centeron{\hbox{$\,#1$}}{\lower 2.5ex\hbox{$\char'176$}}}
\def\sumtil{\centeron{\hbox{$\displaystyle\sum$}}{\lower
-1.5ex\hbox{$\widetilde{\phantom{xx}}$}}}

\def\q{\unc q}

\newcommand{\bit}{\begin{itemize}}
\newcommand{\eit}{\end{itemize}}

\newcommand{\beq}{\begin{equation}}
\newcommand{\eeq}{\end{equation}}
\newcommand{\beqa}{\begin{eqnarray}}
\newcommand{\eeqa}{\end{eqnarray}}

\begin{document} 
\begin{titlepage} 

\rightline{\vbox{\halign{&#\hfil\cr
&ANL-HEP-PR-99-108\cr
&\today\cr}}} 
\vspace{0.25in} 

\begin{center} 
 
{\large\bf 
THE TRIPLE-REGGE TRIANGLE ANOMALY }\footnote{Work 
supported by the U.S.
Department of Energy, Division of High Energy Physics, \newline Contracts
W-31-109-ENG-38 and DEFG05-86-ER-40272} 
\medskip

Alan. R. White\footnote{arw@hep.anl.gov }

\vskip 0.6cm

\centerline{High Energy Physics Division}
\centerline{Argonne National Laboratory}
\centerline{9700 South Cass, Il 60439, USA.}
\vspace{0.5cm}

\end{center}

\begin{abstract} 

It is shown that the triangle anomaly is present, as an infra-red
divergence, in the six-reggeon triple-regge interaction vertex obtained from
a maximally non-planar Feynman diagram in the full triple-regge limit of
three-to-three quark scattering. A multi-regge asymptotic
dispersion relation formalism can be used to isolate all anomaly
contributions and to discuss when and how there is a cancelation.

\end{abstract}

\renewcommand{\thefootnote}{\arabic{footnote}} \end{titlepage}

\mainhead{1. INTRODUCTION}

In a companion paper to this\cite{arw99}
we demonstrate that in massless QCD certain
reggeized gluon interactions contain an infra-red divergence that can be
understood as the infra-red appearance\cite{cg} of the U(1) quark anomaly.
Such vertices appear in a wide variety of multi-regge reggeon
diagrams\cite{arw98} and so this is a potential new manifestation of the
anomaly in a dynamical role. Indeed we believe the appearance of the anomaly 
in this context is crucial for obtaining a unitary pomeron, together with
confinement and chiral symmetry breaking, in (multi-)regge limits. 

It is well-established that regge limits within QCD are described by 
reggeon diagrams\cite{fkl}. Such diagrams involve reggeized gluon (or quark)
propagators,
reggeon interaction vertices, and external couplings to the scattering states,
all of which are gauge invariant. In general, many Feynman diagrams
give contributions to a single reggeon vertex - the BFKL kernel is a
well-known example\cite{fkl}. In this paper we illustrate the results of
\cite{arw99} by presenting an abbreviated version of the central calculation.
We study the full triple-regge limit\cite{gw} of three-to-three quark
scattering and show that the anomaly is present in the (six-reggeon)
triple-regge interaction vertex obtained from a ``maximally non-planar''
Feynman diagram. In \cite{arw99} we use an asymptotic dispersion relation
formalism\cite{arw1,sw} to show that in lowest-order the anomaly occurs only in
such diagrams and in those related to them by reggeon Ward identities. Within
this formalism, the anomaly appears because unphysical multiple discontinuities
containing the necessary chirality transition contribute to the dispersion
relation. For direct Feynman diagram calculations, it's appearance can be
understood as due to an unphysical singular configuration approaching the
asymptotic region in which every propagator in a quark loop is on-shell and
one propagator carries the zero momentum necessary for a chirality transition.

Even though the anomaly is present in a reggeon interaction it does not 
necessarily produce an effect in the amplitudes in which it is contained.
Both the structure of external couplings and additional symmetries of the full
reggeon diagrams can produce a cancelation.
In particular, we show in \cite{arw99} that when the scattering 
states are elementary quarks or gluons the anomaly always cancels in the full 
scattering amplitude. Indeed, in the lowest-order diagrams the symmetries of the
transverse momentum integrations directly produce the cancelation. In general
we expect such cancelations to extend to any process where the chirality 
violation involved can not be linked to
(reggeized) gluon configurations with the quantum numbers of the winding-number
current. 

The anomaly does not cancel when 
a ``reggeon condensate'' with the quantum numbers of 
the winding-number current is present. It was argued 
in \cite{arw98} that, in 
a color superconducting phase with the gauge symmetry broken from SU(3) to 
SU(2), such a condensate is consistently 
reproduced in all reggeon states by anomaly infra-red 
divergences, while also producing confinement, chiral 
symmetry breaking and a regge pole pomeron. SU(3) gauge invariance should be 
obtained by critical pomeron behavior\cite{cri} in which the condensate and 
superconductivity simultaneously disappear. 
In \cite{arw98}
we assumed the existence of the anomaly together with a number of 
properties that the results of \cite{arw99} show to be essentially correct, 
although there are significant differences. Once we
have determined the full structure of the anomaly, 
we hope to implement the program of \cite{arw98} in detail in future papers. 
If a unitary (reggeon) S-Matrix is obtained, as we anticipate, it will
be very close to perturbation 
theory, with the non-perturbative properties of confinement and chiral 
symmetry breaking a consequence of the anomaly only. 

We study the three-to-three scattering process illustrated in Fig.~1(a)
and define momentum transfers $Q_1, Q_2$ and $Q_3$ as in Fig.~1(b).
\begin{center}
\leavevmode 
\epsfxsize=2.7in
\epsffile{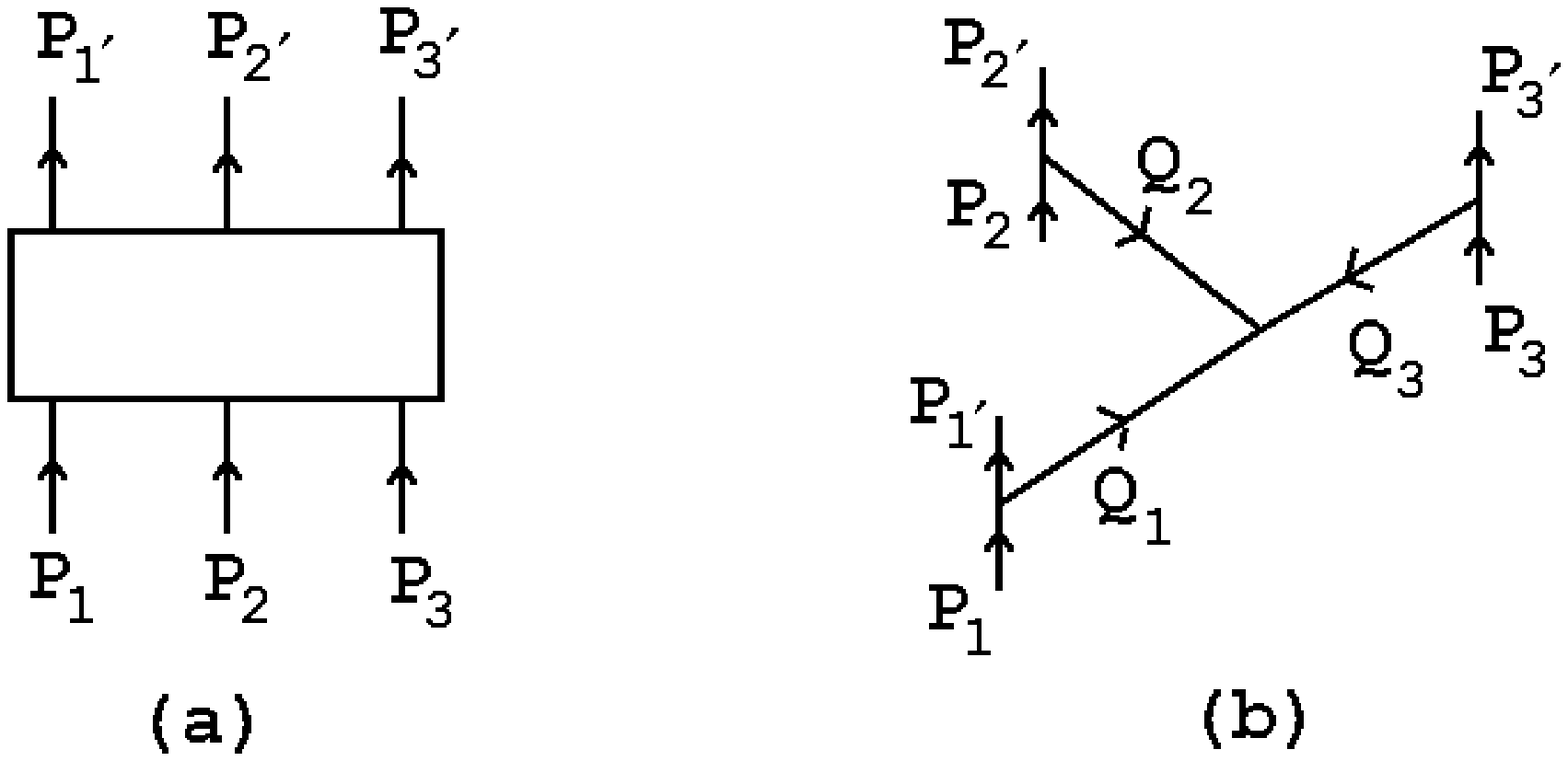}

Fig.~1 Three-to-Three Scattering.

\end{center}
The full triple-regge limit\cite{gw} can be realized by taking 
each of $P_1,~P_2$ and $P_3$ large 
along distinct light-cones, i.e.
\newline \parbox{3in}{ 
$$
\eqalign{ P_1~\to&~ P_1^+~= ~(p_1,p_1,0,0)~,~~p_1 \to \infty \cr
P_2~\to&~ P_2^+~= ~(p_2,0,p_2,0)~,~~p_2 \to \infty \cr
P_3~\to&~ P_3^+~= ~(p_3,0,0,p_3)~,~~p_3 \to \infty  }
$$}
\parbox{3in}{
$$ \eqalign{
~~~Q_1~\to&~~ (\hat{q}_1,\hat{q}_1,q_{12},q_{13})\cr
~~~Q_2~\to&~ ~(\hat{q}_2,q_{21},\hat{q}_2,q_{23})\cr
~~~Q_3~\to&~~(\hat{q}_3,q_{31},q_{32},\hat{q}_3)}
\auto\label{np3}
$$}
Momentum conservation requires that 
$$
\hat{q}_1 + \hat{q}_2 + \hat{q}_3 = 0,~ \hat{q}_1 + q_{21} 
+q_{31}=0,~ \hat{q}_2 + q_{12} 
+q_{32}=0,~ \hat{q}_3 + q_{13} 
+q_{23}=0
\auto
$$
and so there are a total of five independent $q$ variables which, along  
with $p_1, p_2$ and $p_3$, give the necessary eight variables. 

Consider the maximally non-planar Feynman diagram shown in Fig.~2.
In the limit (\ref{np3}), leading behavior is obtained when the quark loop 
remains at rest and the gluons are exchanged between the fast quarks and the 
loop. Some combination of quark propagators will also be placed on  
(or close to) mass-shell by the 
limit. For the even signature amplitudes that we are interested in
only on-shell (and not close to on-shell) propagators contribute and
all gluons in the
diagram are the lowest-order contribution of a reggeized gluon. 
\begin{center}
\leavevmode 
\epsfxsize=4in
\epsffile{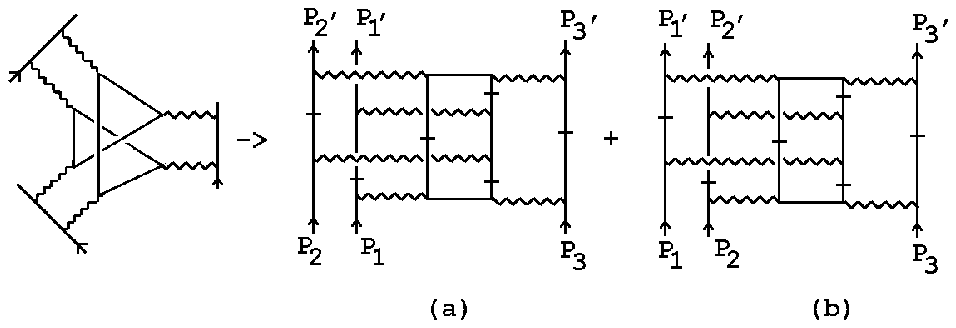}

Fig.~2 Contributions from a Maximally Non-Planar Diagram

\end{center} 

We discuss below how we determine whether a particular combination of on-shell
propagators gives a relevant contribution. For the moment, we note only that
Fig.~2(a) and (b) represent two possibilities. If we consider time to
be in the vertical direction on the page then, a-priori, the hatched quark
lines can naturally be close to mass-shell during the scattering. (Although,
since the anomaly involves an unphysical chirality transition, 
we do  not expect to find it 
in a succession of on-shell physical scattering process of this kind.)
If we sum over quark directions, the loop 
lines hatched in Fig.~2(a) are the unhatched lines in Fig.~2(b). In either
case, if we put all hatched lines on-shell and use the corresponding
$\delta$-functions to carry out longitudinal integrations in each 
of the external loops, we obtain a contribution to the lowest-order
six-reggeon interaction as follows. 
We write $q_i=Q_i/2,~i=1,2,3 $~ and, for Fig.~2(a), 
label the gluon momenta as shown in Fig.~3(a).
As is also shown in this figure, in the limit
(\ref{np3}) the gluons couple to the quark loop via a 
``light-like'' $\gamma$-matrix determined by the external momenta, 
i.e. $P_i^+ \to~ \gamma_{i^-} = \gamma_0 -\gamma_i~, i=1,2,3$. 
\begin{center}
\leavevmode 
\epsfxsize=4in
\epsffile{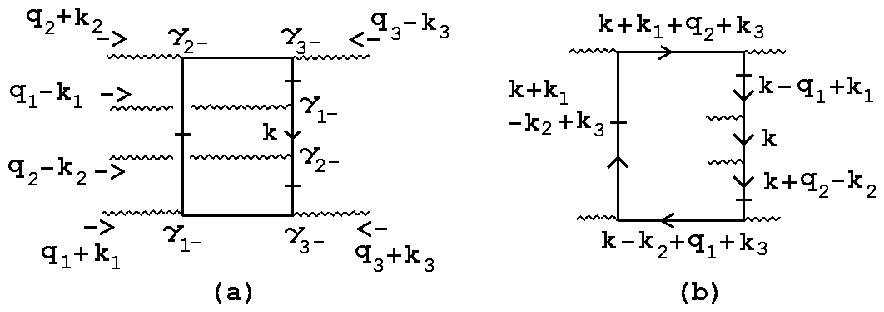}

Fig.~3 (a) Gluon Momenta and $\gamma$-Matrix Couplings (b) Quark Loop Momenta 
\end{center} 

To carry out the longitudinal $k_3$ integrations we use 
conventional light-cone co-ordinates. For the $k_1$ and $k_2$ 
integrations we use special light-cone co-ordinates\cite{arw99}, i.e.
for a general four-momentum $p^{\mu}=(p_0,p_1,p_2,p_3)$ we write  
$$
p^{\mu}  = p_{2^-} \underline{n}_{1^+}~ +~ p_{1^-}  
\underline{n}_{2^+} ~ + ~ 
\underline{p}_{12+}~, 
\auto\label{lcd4}
$$
where $\underline{p}_{12+}= p_{12-}\underline{n}_{12+} 
~+~ p_3\underline{n}_3~$ with $ p_{12-} = p_1 + p_2 - p_0 ~$, and
$$
\underline{n}_{1^+} = (1,1,0,0)~,~~\underline{n}_{2^+} = (1,0,1,0)~,~~
\underline{n}_{12+} = (1,1,1,0)~,~~ \underline{n}_{3} = (0,0,0,1)
\auto\label{unvs}
$$
Analagous decompositions to (\ref{lcd4}) can also be introduced for 
$\gamma$-matrices. 

With the hatched lines on-shell the quark loop reduces to a 
triangle diagram. With the light-cone co-ordinates we have chosen 
only the momentum routing of Fig.~3(b) produces the local (momentum independent)
couplings at all
three vertices\cite{arw99} which are a prerequisite
for the anomaly to be present. 
Apart from a normalization and color factor 
the asymptotic amplitude obtained is 
$$
\eqalign{& ~~~~~~~~~~~~~ g^{12} ~~{p_{1}~p_{2}~p_{3} \over m^3} 
~~~ \times \cr 
&\int { d^2 \underline{k}_{112+} \over
(q_1 + \underline{k}_{112+})^2(q_1 - \underline{k}_{112+})^2}
~\int {d^2  \underline{k}_{212} \over
(q_2 + \underline{k}_{212+})^2(q_2 - \underline{k}_{212+})^2}
~\int  {d^2  \underline{k}_{33\perp} \over
(q_3 + \underline{k}_{33\perp})^2(q_3 - \underline{k}_{33\perp})^2} 
\cr  
&\int d^4 k~{ Tr \{ \hat{\gamma}_{12} (\st{k}+ \st{k}_1 
+ \st{q}_2 + \st{k}_3 +m) 
\hat{\gamma}_{31}( \st{k} +m) 
\hat{\gamma}_{23} (\st{k}- \st{k}_2 + \st{q}_1 
+ \st{k}_3 +m) \} \over 
([k + k_1 + q_2 + k_3]^2 - m^2) 
(k^2 - m^2) 
 ([k - k_2 + q_1 +k_3]^2 - m^2)} ~+ ~ \cdots }
\auto\label{578}
$$
where $k_{11^-}= k_{22^-} = k_{33^-}=0$, 
$~k_{12^-}, k_{21^-}$ and $k_{33^+}$ 
are determined by the mass-shell constraints that put propagators on-shell,
and 
$$
\eqalign{ \hat{\gamma}_{31} & ~=~\gamma_{3^-} \gamma_{2^-} \gamma_{1^-} 
~=~ \gamma^{-,+,-}~- ~i~\gamma_5 \gamma^{-,-,-} \cr
~\hat{\gamma}_{23} &~=~\gamma_{2^-} \gamma_{1^-} \gamma_{3^-} 
~=~  \gamma^{+,-,-}~-~ i~ \gamma_5 \gamma^{-,-,-} \cr
\hat{\gamma}_{12} &~=~ \gamma_{1^-} \gamma_{3^+} \gamma_{2^-} ~=~
\gamma^{-,-,-} ~+ ~i ~\gamma_5 \gamma^{-,-,+}  }
\auto\label{576}
$$
with 
$$ 
\gamma^{\pm,\pm,\pm} ~=~ \gamma^{\mu}\cdot n^{\pm,\pm,\pm}_{ \mu} ~,~~~~~~
n^{\pm,\pm,\pm \mu} ~= ~ (1,\pm1,\pm1,\pm1) 
\auto\label{5760}
$$
In that part of the amplitude not shown explicitly in (\ref{578}) 
non-local (momentum dependent)
couplings replace one, or more, of the $\hat{\gamma}$'s. 

Writing $~\alpha_i = \alpha((q_i+k_i)^2)+ \alpha((q_i-k_i)^2)-1, ~i=1,2,3~,$
where $\alpha(t)= 1 +~O(g^2)$ is the gluon regge trajectory, we have
$$
\eqalign{~~~~~~~p_{1}~& p_{2}~ p_{3}~=~(p_1p_2)^{{1\over 2}}
(p_2p_3)^{{1\over 2}}(p_3p_1)^{{1\over 2}}~
\sim ~s_{12}^{1/2}~s_{23}^{1/2}~
s_{31}^{1/2}~\cr
&= ~(s_{13})^{(\alpha_1+\alpha_3-\alpha_2)/2}
(s_{23})^{(\alpha_2+\alpha_3-\alpha_1)/2}
(s_{12})^{(\alpha_1+\alpha_2 -\alpha_3)/2} ~~+~ O(g^2) }
\auto\label{212}
$$
where $s_{ij}=(p_i + p_j)^2$. This is the 
lowest-order triple-regge behavior 
for the amplitudes that interest us (and, in particular, potentially contain
the anomaly). 
The transverse momentum integrals and gluon propagators in (\ref{578})
can then be interpreted\cite{arw99,arw98} as the lowest-order 
contributions of two-reggeon states in each $t_i ~(=Q_i^2)$-channel.
Removing these factors, the three $\gamma_5$ couplings in (\ref{576}) give 
the $m=0$ reggeon interaction 
$$
\eqalign{ &\Gamma_6(q_1,q_2,q_3,
\tilde{\underline{k}}_1,\tilde{\underline{k}}_2, 
\underline{k}_{3\perp},0) ~=\cr
& \int d^4 k  {  Tr \{ 
\gamma_5 \gamma_{-,-,+} (\st{k}+ \st{k}_1 + \st{q}_2 +\st{k}_3) 
\gamma_5 \gamma_{-,-,-} ~\st{k}~ 
\gamma_5 \gamma_{-,-,-}(\st{k}- \st{k}_2 + \st{q}_1 + \st{k}_3 ) 
\over  (k + k_1 + q_2 + k_3 )^2  
~k^2 ~
 (k - k_2 + q_1 + k_3)^2 }  ~+ ~ \cdots }
\auto\label{580}
$$
corresponding to the triangle diagram of Fig.~4. 
\begin{center}
\leavevmode
\epsfxsize=1.9in
\epsffile{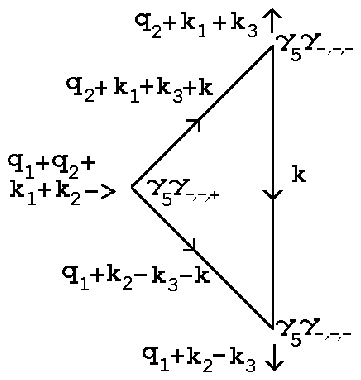}

Fig.~4 The Triangle Diagram Corresponding to (\ref{580})
\end{center}

The maximal anomaly infra-red divergence\cite{arw99,cg} 
of the axial-vector triangle diagram
tensor  $\Gamma^{\mu\nu\lambda}$  occurs in components with
$\mu= \nu $ lightlike and $\lambda $ 
an orthogonal spacelike index. In our case $\gamma_{-,-,-}$ appears at 
two vertices and this has a projection on the 
light-like vector 
$\underline{n}_{lc}= (1, cos {\theta}_{lc}, sin {\theta}_{lc}, 0)$.
The $\gamma_{-,-,+}$ factor at the third vertex has a distinct projection on the
orthogonal $\underline{n}_3$ direction. 
The anomaly produces a linear divergence\cite{arw99,cg} when we take the limit
$$
(k_1 + q_2 +k_3)^2 \sim (q_1 + q_2 + 
k_1 +k_2)^2 \sim (k_2 + q_1 - k_3)^2 \sim 
\hbox{\q}^2 \to 0
\auto\label{5801}
$$
of (\ref{580}) while keeping a finite light-like momentum, parallel to 
$\underline{n}_{lc}$, flowing through the diagram. 
To show that this can be 
done with the mass-shell constraints satisfied,  we define $
\underline{n}_{lc\perp}= (0, - sin {\theta}_{lc},cos {\theta}_{lc},  0)
$ 
orthogonal to $\underline{n}_{lc}$ and take 
$$
\eqalign {q_1 + k_1 + q_2 + k_2~&=~O(\hbox{\q})~\underline{n}_{lc\perp} \cr
k_3 -k_2 -q_1~=~ q_2 -k_2+q_3 +k_3 ~&= ~l~\underline{n}_{lc} ~+~
O(\hbox{\q})~\underline{n}_{lc\perp} \cr
q_2 +k_1+k_3~=~ k_1 -q_1+k_3 -q_3~&= ~ l~\underline{n}_{lc} ~+~
O(\hbox{\q})~\underline{n}_{lc\perp} }
\auto\label{250}
$$
We also take the loop momentum $k \sim O(\hbox{\q})$ 
and let $\hbox{\q} \to 0 $ with 
$$
(q_1 - k_1) ~\to~-2 l~(1,1,0,0)~, ~~~~~(q_2- k_2) ~\to~2 l~(1,0,1,0)~,
\auto\label{2510}
$$
and 
$$
q_3 ~\to ~l~(0,1-1,0)~, ~~~~~k_3~\to~l~(0,1-2cos {\theta}_{lc}~, 
1-2 sin {\theta}_{lc}~, 0)
\auto\label{2511}
$$ 
In the limiting configuration 
the hatched lines 
of Fig.~2(a) are on-shell and in addition to $~(q_1+k_1)^2=(q_2+k_2)^2~$
we also have $~q_1^2=q_2^2= k_1^2 =k_2^2~ $.
Only the lightlike momentum $k_{lc}=l~ \underline{n}_{lc}$
flows through the triangle graph of Fig.~4. The anomaly 
divergence appears, therefore, and the projection on $\underline{n}_{lc}$ gives
$$
\Gamma_6 ~~\sim  ~~ {(1 - cos {\theta}_{lc} - sin {\theta}_{lc})^2
~l^2 \over \hbox{\q} } 
\auto\label{5847}
$$

It might appear that we have found the anomaly with the quark lines on-shell as 
a consequence of the physical scattering illustrated in 
Fig.~2(a). However, it is not difficult to show\cite{arw99} that we have chosen
what would be the unphysical pole (in this scattering) for the left-side
propagator in the quark loop. In fact, 
the momentum configuration (\ref{2510}) and (\ref{2511}) actually describes the 
physical scattering illustrated in Fig.~5(a), if the time axis is 
vertical on the page. 
\begin{center}
\leavevmode
\epsfxsize=4in
\epsffile{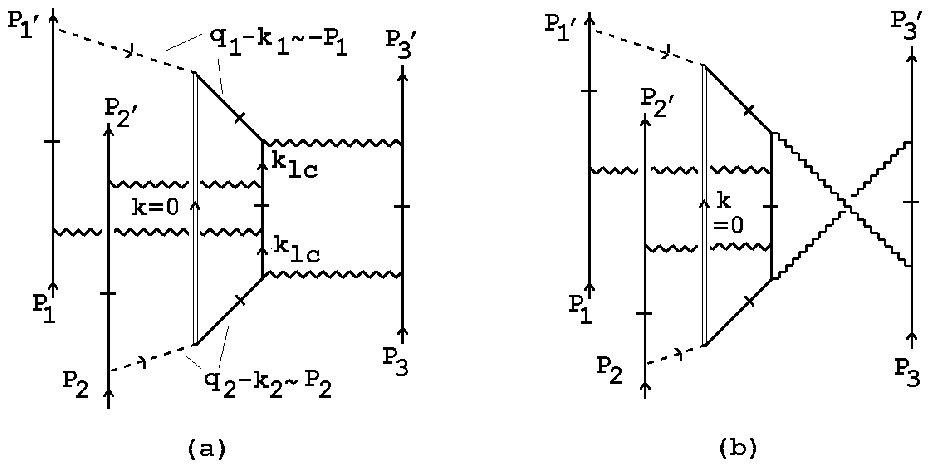}

Fig.~5 Physical Scattering Processes Involving the Anomaly
\end{center}
This is the basic process associated with 
the anomaly in the reggeon vertex obtained from the 
lowest-order graphs. The dashed lines indicate light-like (``wee
parton'') gluons, one incoming produced by an incoming quark and one outgoing
that is absorbed by an outgoing quark. 
A zero-momentum quark (indicated by the open line) is emitted
by the incoming wee-parton gluon, undergoes a chirality transition, and then 
is absorbed by the outgoing wee-parton gluon. The accompanying (on-shell)
antiquark has
it's incoming light-like momentum pointed along $\underline{n}_{lc}$ by
scattering off a spacelike gluon. It then forward scatters off two more
spacelike gluons before another
scattering points it's lightlike momentum in the outgoing direction.

Fig.~5(a) is quite distinct from the implicit time-ordering of 
Fig.~2(a) that we anticipated might place the hatched lines on-shell.
In fact, in Fig.~5(a) the initial and final light-like anti-quarks are clearly
not part of a physical intermediate state. Rather, the on-shell configurations
enter the physical region only as the wee gluons become light-like. 
In the multi-regge dispersion relation formalism\cite{arw99},
that we briefly elaborate on below, 
all on-shell propagators are associated with some contributing discontinuity.
If this discontinuity is not physical it 
must necessarily approach the physical region 
asymptotically, either from a cross-channel physical region as part of
a physical multiple discontinuity (such discontinuities can not contain
chirality transitions) or from an unphysical direction as part 
of an unphysical multiple discontinuity. For Fig.~5(a) 
the initial and final quark/antiquark
pair can have the net chirality due to effective axial couplings 
only if the on-shell hatched lines represent unphysical discontinuities.

Before discussing the dispersion relation formalism we first consider
diagrammatically how the anomaly might be canceled. 
Diagrams that contain the same quark loop divergence must be involved and
an immediate possibility is the diagram and space-time scattering of
Fig.~5(b). In an abelian theory Ward identities relating the
diagrams of Fig.~5(a) and (b) at $(q_i \pm k_i)^2 =0,~i=1,2$ would 
produce a cancelation of the anomaly. With color factors, however, the two
wee-parton gluons can form a color octet state in which the two
diagrams do not cancel. The corresponding non-abelian (reggeon)
Ward identities involve a triple gluon diagram that can not produce the
chirality transition of the anomaly. Indeed, the failure of the reggeon
Ward-identities can be regarded as the defining property that indicates the
presence of the anomaly in the reggeon vertex.

There is, however, a relatively simple cancelation that occurs 
within the complete Feynman diagram (or more generally reggeon diagram) 
that contains the reggeon vertex. 
When the hatched lines of Fig.~2(b) are placed on-shell, the amplitude
obtained differs only from that given by Fig.~2(a) in that the roles of $P_1$
and $P_2$ are interchanged, together with $k_1 \to -k_1$ and $k_2 \to -k_2$. 
In Fig.~5(a) the incoming and outgoing wee parton
gluons are correspondingly interchanged.
This interchange can be viewed as a ($t$-channel) parity transformation
that produces a change of sign of the anomaly. After transverse momentum
integration, the rest of the 
diagram is kinematically insensitive to this
transformation and the anomaly is necessarily canceled.

In higher-orders the two gluons in each $t_i$ channel are replaced by
even signature two-reggeon states.
If we keep the same quark loop interaction,
the amplitude with even signature in each channel is obtained 
by summing over diagrams related by a twist in each $t_i$-channel.
For external quark scattering, both the color factors and all three
of the $k_i$-integrations are then symmetric and the anomaly cancels.
Sufficient antisymmetry to avoid cancelation would appear only if each of the
even signature two-reggeon states carried negative color parity. However,
anomalous color parity (i.e. $\neq$ signature) reggeon states do not couple to
elementary quarks (or gluons),

So far we have discussed only specific contributions from selected diagrams. 
In higher-orders, as both quarks 
and gluons reggeize, we expect the maximal non-planarity 
property to be necessary to produce, in each regge channel, the double spectral 
function property that is well-known to be required for regge cut couplings.
In lowest-order we can not appeal to this expectation and, a-priori,
all diagrams containing a quark loop and two gluons exchanged in each $t_i$
channel could contribute. 
The total number of diagrams of this form is $O(100)$ and so studying them 
individually would clearly be difficult, if not impossible.

Fortunately, it is possible to systematically count all
anomaly contributions by using the multi-regge asymptotic dispersion relation
formalism developed in \cite{arw1} and \cite{sw}. This is described in detail
in \cite{arw99} and here we give only a very brief discussion. The
essential feature is that the asymptotic cut structure of amplitudes reduces
to normal threshold branch-cuts satisfying the Steinmann relations, i.e. no
double discontinuities in overlapping channels. Consequently the full
amplitude can be written as
$$
M(P_1,P_2,P_3,Q_1,Q_2,Q_3)~ =~ 
\sum_{\cal C} M^{\cal C}(P_1,P_2,P_3,Q_1,Q_2,Q_3)
~+~M^0~,\auto\label{dis}
$$
where $M^0$ contains all non-leading triple-regge behavior, double-regge
behavior, etc. and the sum is over all triplets ${\cal C}$ of three
non-overlapping, asymptotically distinct, cuts allowed by the Steinmann
relations.

When formulated in terms of angular variables the triple-regge limit is 
described as $z_i \to \infty~, i=1,2,3$, where $z_i= cos \theta_i$ and 
$\theta_i$ is a $t_i$-channel scattering angle. 
If we consider only the $z_i$ dependence, the asymptotic behavior of
invariants is 
\beqa
s_{122'} ~&\sim&~ s_{1'3'3}~\sim ~- s_{1'22'} ~\sim~ - s_{13'3} 
~\sim ~~z_1 \nonumber \\
s_{233'} ~&\sim&~ s_{2'1'1}~\sim ~- s_{2'33'} ~\sim~ - s_{21'1} 
~\sim ~~ z_2 \nonumber \\
s_{311'} ~&\sim&~ s_{3'2'2}~\sim ~- s_{3'11'} ~\sim~- s_{32'2} 
~\sim ~~ z_3 \nonumber \\
s_{13} ~&\sim&~ s_{1'3'}~\sim~ -s_{13'}~\sim~ -s_{1'3} 
~\sim ~~ z_1~ z_3 \nonumber \\
s_{23} ~&\sim&~ s_{2'3'}~\sim~ -s_{23'}~\sim ~-s_{2'3} 
~\sim ~~ z_2~ z_3~ \nonumber \\
s_{12} ~&\sim&~ s_{1'2'} ~\sim~ -s_{12'}~ \sim~ -s_{1'2} 
~\sim ~~ z_1~ z_2~  \label{zap}  
\eeqa
where, as above, $s_{ij}=(P_i+P_j)^2$, and in addition 
$s_{ij'}=(P_i-P_{j'})^2$, $s_{ijj'}=(P_i+P_j-P_{j'})^2$, etc.  
From (\ref{zap}) we find that all the asymptotic triple discontinuities are of
one of the three forms, illustrated in Fig.~6 by tree diagrams
in which an internal line represents a channel discontinuity. 
There are 24 of the first kind shown
and 12 each of the second and third kinds. When the three invariants 
involved in an (a) or 
\begin{center}
\leavevmode
\epsfxsize=2.5in
\epsffile{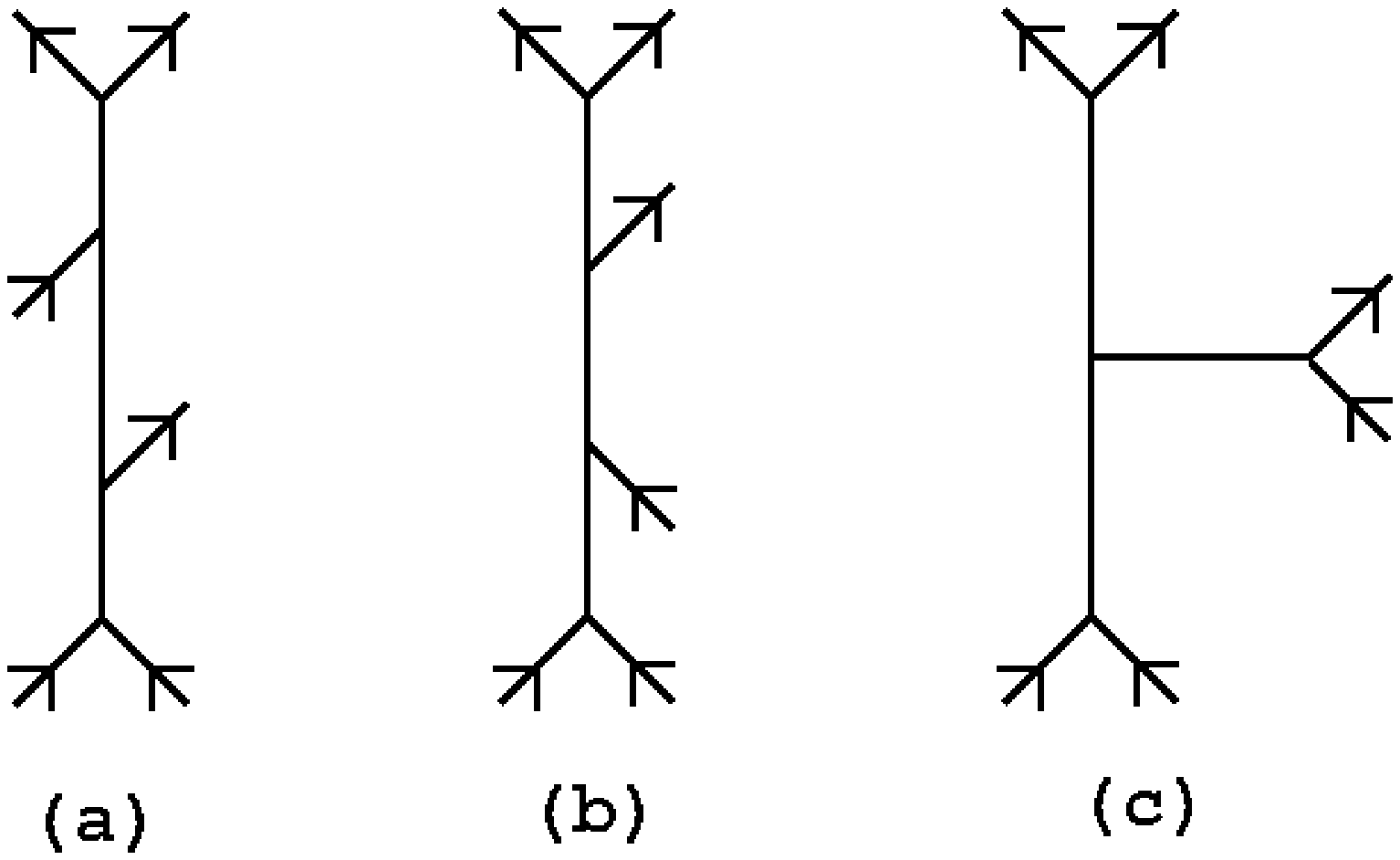}

Fig.~6 Tree Diagrams for Triple Discontinuities.
\end{center}
(b) kind are each large and 
positive the $z_i$ are always 
real and, as a result, such triple discontinuities lie in physical regions
and include only 
on-shell configurations of the kind illustrated in Figs.~2(a) and 
(b). Crucially, however, asymptotic triple discontinuities of the (c) kind occur
only in unphysical regions where the $z_i$ are pure imaginary and some
invariants (not giving a discontinuity) are also pure imaginary. Such
unphysical discontinuities do not appear in simpler multi-regge asymptotic
dispersion relations.

Unphysical discontinuities can not be discussed directly from
an S-Matrix starting-point.
However, the asymptotic dispersion relation can also be obtained\cite{arw00}
by starting with spacelike masses and utilising the primitive analyticity 
domains that follow from the field theory formalism of Generalised Retarded
Functions\cite{cs}. The unphysical triple discontinuities 
appear 
just because they satisfy the Steinmann
relations. These discontinuities are physical in two-four
scattering processes and, in principle at least, S-Matrix ``extended
unitarity'' equations can be derived from each of the three neighboring
two-four regions for the three-three processes
we discuss. It is then obvious that multiple discontinuities will be present
combining processes from separate physical regions and in 
which chirality transitions can naturally occur.

To avoid introducing multi-regge theory in this paper we
simply note that with the use of Sommerfeld-Watson representations 
and (the assumption of) regge behavior, the multiple discontinuities in the
asymptotic dispersion relation can be converted to amplitudes that 
analytically continue away from the discontinuity region.
To study the full triple-regge amplitude, therefore, it suffices to calculate 
all asymptotic triple discontinuities. Since there are only 
three distinct kinds, this is a much simpler proposition than the task
of studying separately all Feynman diagrams.

The multi-regge representations for Fig.~6(b) and (c) are significantly 
different from that for Fig.~6(a). 
In particular, while the Fig.~6(a) discontinuities produce 
amplitudes with three possible signatures, each of the 
sets of Fig.~6(b) and (c) provide 
only four distinct signatured amplitudes. (We anticipate that the resulting  
``signature conservation'' rule will lead to
the even signature property of 
the pomeron when we finally extract the physical S-Matrix from reggeon 
diagrams.) Also the resulting kinematic structure implies 
that the anomaly can appear only in the (b) or (c) types. 
The need for an unphysical chirality transition then selects the
(c) type and it is possible to show that the anomaly arises only from
discontinuities of this type and only from diagrams of the kind 
we have discussed. The above analysis can then be 
recast as the calculation of an unphysical multiple
discontinuity followed by an extrapolation that gives 
the physical region anomaly amplitude of Fig.~5(a). 
The arguments for cancelation 
in the scattering of elementary quarks or gluons will apply.


\begin{thebibliography}{99}

\bibitem{arw99} A.~R.~White,  hep-ph/9910458 (ANL-HEP-PR-99-102).

\bibitem{cg} S.~Coleman and B.~Grossman, {\it Nucl. Phys. }
{\bf B203}, 205 (1982).

\bibitem{arw98} A.~R.~White, {\it Phys. Rev.} {\bf D58}, 074008 (1998), see 
also Lectures in the Proceedings of the Theory Institute on Deep-Inelastic
Diffraction, Argonne National Laboratory (1998).

\bibitem{fkl} E.~A.~Kuraev, L.~N.~Lipatov, V.~S.~Fadin, {\it Sov. Phys.
JETP} {\bf 45}, 199 (1977);
J.~B.~Bronzan and R.~L.~Sugar, {\it Phys. Rev.} {\bf D17}, 
585 (1978), his paper organizes into reggeon diagrams the results from 
H.~Cheng and C.~Y.~Lo, Phys. Rev. {\bf D13}, 1131 (1976), 
{\bf D15}, 2959 (1977); 
V.~S.~Fadin and V.~E.~Sherman, Sov. Phys. JETP {\bf 45}, 
861 (1978);
V.~S.~Fadin and L.~N.~Lipatov, {\it Nucl. Phys.} {\bf B477},
767 (1996) and further references therein;
J.~Bartels, {\it Z. Phys.} {\bf C60}, 471 (1993) and further
references therein; A.~R.~White, 
{\it Int. J. Mod. Phys.} {\bf A8}, 4755 (1993). 

\bibitem{gw} P.~Goddard and A.~R.~White, {\it Nucl. Phys.} {\bf B17}, 1, 45
(1970).

\bibitem{arw1} A.~R.~White, Int. J. Mod. Phys. {\bf A11}, 1859 (1991);
 A.~R.~White in {\em Structural Analysis of Collision Amplitudes},
proceedings of the Les Houches Institute, eds. R.~Balian and D.~Iagolnitzer 
(North Holland, 1976); H.~P.~Stapp ibid.

\bibitem{sw} H.~P.~Stapp and A.~R.~White, {\it Phys. Rev.} {\bf D26}, 2145 
(1982). 

\bibitem{cri}  A.~A.~Migdal, A.~M.~Polyakov and K.~A.~Ter-Martirosyan, 
{\it Zh. Eksp. Teor.  Fiz.} {\bf 67}, 84 (1974); 
H.~D.~I.~Abarbanel and J.~B.~Bronzan, {\it Phys. Rev.} {\bf D9}, 2397 (1974).

\bibitem{arw00} A.~R.~White, ``The Past and Future of S-Matrix Theory'',
hep-ph/0002303 (ANL-HEP-PR-00-011), to be published in ``Scattering'', edited
by E.~R.~Pike and P.~Sabatier.

\bibitem{cs} K.~E.~Cahill and H.~P.~Stapp {\it Ann. Phys.} {\bf 90}, 438
(1975).



\end{thebibliography}
\end{document}